\documentclass[prb,twocolumn,superscriptaddress,showpacs]{revtex4}
\usepackage{graphicx,amsmath,psfrag,color,amssymb}

\usepackage{ulem}
\begin{document}

\title{Quantum quench dynamics and population inversion in bilayer graphene} 

\author{Bal\'azs D\'ora}
\email{dora@kapica.phy.bme.hu}
\affiliation{Max-Planck-Institut f\"ur Physik komplexer Systeme, N\"othnitzer
  Str. 38, 01187 Dresden, Germany} 
\affiliation{Department of Physics, Budapest University of Technology and
  Economics, Budafoki \'ut 8, 1111 Budapest, Hungary} 

\author{Eduardo V. Castro}
\affiliation{Instituto de Ciencia de Materiales de Madrid, CSIC, Cantoblanco,
  E-28049 Madrid, Spain} 
\affiliation{Centro de F\'isica do Porto, Rua do Campo Alegre 687, P-4169-007 
  Porto, Portugal} 
                                   
\author{Roderich Moessner} 
\affiliation{Max-Planck-Institut f\"ur Physik komplexer Systeme, N\"othnitzer
  Str. 38, 01187 Dresden, Germany} 

\date{\today} 

\begin{abstract}
The gap in bilayer graphene (BLG) can directly be controlled by a perpendicular
electric field. By tuning the field through zero at a finite rate in neutral BLG,
excited states are produced. Due to screening, the resulting dynamics is
determined by coupled non-linear Landau-Zener models. The generated defect
density agrees with Kibble-Zurek theory in the presence of subleading
logarithmic corrections. After the quench, population inversion occurs for
wavevectors close to the Dirac point. This could, at least in principle, provide a coherent
source of infra-red radiation with tunable spectral properties (frequency and
broadening). Cold atoms with quadratic band crossing exhibit the same
dynamics.
\end{abstract}

\pacs{81.05.Uw,64.60.Ht,78.67.-n}
                                   
\maketitle

\section{Introduction}

Charge carriers in bilayer graphene (BLG), which 
consists of two atomic layers of crystalline carbon,  
combine non-relativistic "Schr\" odinger" (quadratic dispersion) and 
relativistic "Dirac" (chiral symmetry, unusual Berry phase) features.
Due to their peculiar nature, BLG holds the promise of 
revolutionizing electronics, since its band gap is directly controllable 
by a perpendicular electric field over a wide range of 
parameters~\cite{mccannscreening,evcastro,oostinga,MLS+09,XFL+10} 
(up to 250~meV~\cite{ZTG+09}), unlike existing 
semiconductor technology. Moreover, unlike  monolayer graphene (MLG), whose
effective model (the Dirac equation) was thoroughly studied in QED and relativistic
quantum mechanics, understanding the  low energy properties of BLG is a new
challenge.

Tuning the gap through zero in BLG in a time dependent perpendicular electric field parallels closely to
 a finite rate passage through a quantum  critical point (QCP): 
as the gap closes, activated behaviour and a finite correlation length give way to metallic response 
and power-law correlations, as in a sweep through a QCP. During the latter, 
defects (excited states, vortices) are produced according to Kibble-Zurek theory\cite{kibble,zurek}. 
When the relaxation  time of the system, which encodes how much time it 
needs to
adjust to new thermodynamic conditions, becomes comparable to the remaining ramping time  to the critical point,
the system crosses over from  the adiabatic  to the diabatic (impulse) regime. In the latter regime,
its state is effectively frozen, so that it cannot follow the time-dependence of the instantaneous ground states -- as a result,
excitations are produced\cite{damski}.
Evolution restarts only after leaving the diabatic regime, with an initial state mimicking the frozen one.
The theory, general as it is, finds application in very different
contexts in physics, ranging from the early universe cosmological
evolution\cite{kibble} through liquid $^{3,4}$He~\cite{bauerle,zurek,rutuu} and liquid crystals\cite{lc1,lc2}
to ultracold gases\cite{sadler}, verified for both thermodynamic and quantum phase transitions\cite{zoller}.
The relative case of manipulating the gap -- in particular in real time -- 
via a spatially uniform external electric field,  which can therefore play the role of a (time dependent)
control parameter, establishes BLG as an ideal 
setting for the study of quantum quenches with sudden, continuous or any 
other sweep protocols\cite{barankov,sen,grandi}.
This in turn leads to the question: what might such states be useful for?

\begin{figure}[h!]
\includegraphics[width=0.9\columnwidth]{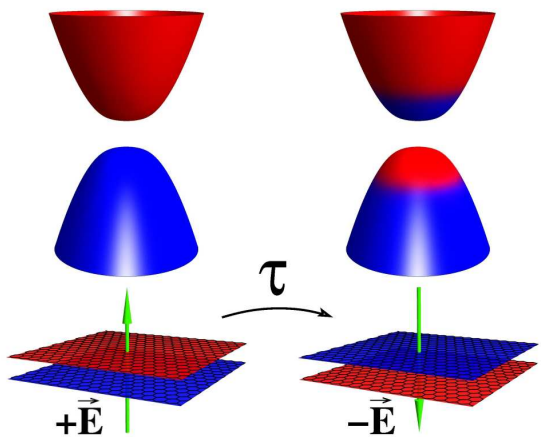}
\caption{Reversing the applied 
perpendicular electric field $+\vec E$ in half-filled BLG (left) at a finite 
rate $1/\tau$ leads to excited states in the upper branch in accordance with 
the Kibble-Zurek theory of non-equilibrium phase transitions (right).
The momentum distribution increases from red  (0) to blue (1) in the spectra.
Realistic quenching times provide an effective population inversion
with little effect on the layer charge asymmetry.
\label{fancy}}
\end{figure}

This complex of questions is addressed here. In particular,
 we compute the defect (excited state) density after a slow, 
non-adiabatic gap-closing passage in BLG via Kibble-Zurek~\cite{kibble,zurek} theory, taking screening between the layers into account.
The presence of excited states after such a quench leads to population 
inversion for wavevectors near the Dirac point in BLG (see Fig.~\ref{fancy}), evidenced by 
the dynamic conductivity. 
This could in principle provide a coherent source of infra-red radiation with tunable spectral properties 
(frequency and broadening), determined below in an idealised model. This is promising as there are only 
few materials that generate light in the infrared with tunable frequency; BLG with its unique properties
might represent the first step towards new lasers for this regime.

\section{Hamiltonian, topological properties}

We study the problem in a more general setting of a general class of low 
energy Hamiltonians, comprising mono- and bilayer graphene, which exhibit
quantum critical behaviour, as
\begin{gather}
H=\left(\begin{array}{cc}
\Delta & c_J(p_x-ip_y)^J\\
c_J(p_x+ip_y)^J & -\Delta
\end{array}
\right),
\label{hamilton}
\end{gather}
where $J$ is a positive integer. The energy spectrum is given by
$E_\pm(p)=\pm\sqrt{\Delta^2+\varepsilon^2(p)}$ with $\varepsilon(p)=c_J|p|^J$
the gapless spectrum,  $|p|=\sqrt{p_x^2+p_y^2}$ with spatial dimension
$d=2$. 

The critical exponents can straightforwardly be read off. 
The correlation length follows from dimensional analysis: 
$\xi\sim\hbar (c_J/|\Delta|)^{1/J}$, defining $\nu=1/J$. 
The Hamiltonian contains the $J$th spatial derivative 
($J$th power of $p$), which leads to $z=J$.
 The resulting scaling relation $z\nu = 1$ is in agreement with a 
linearly vanishing gap $\Delta$.
To understand the nature of this criticality, let us take a closer topological look at Eq. \eqref{hamilton} by evaluating the Berry 
curvature ($\Omega_p$) for a given $J$, which is related to the phase picked up during an adiabatic excursion in the Brillouin zone 
as\cite{berry}
\begin{equation}
\Omega_p=\nabla_{\bf p}\times {\bf }\bf A(\bf p)
\end{equation}
with ${\bf A(p)}=-i\langle n{\bf p}|\nabla_{\bf p}|n{\bf p}\rangle$, $|n{\bf p}\rangle$ is the eigenfunction in the $n$th band.
For Eq. \eqref{hamilton}, we obtain for the $z$ component of the Berry curvature per valley and spin 
\begin{equation}
\Omega_p^z=\frac{\Delta}{2E_+(p)} \left(\frac{{d} \varepsilon(p)}{{d} |p|}\right)^2=
\frac{\Delta J^2c_J^2|p|^{2(J-1)}}{2(\Delta^2+\varepsilon^2(p))^{3/2}},
\label{chern}
\end{equation}
and its integral defines a topological invariant\cite{tkkn} as
\begin{equation}
\mathcal{C}_J=\frac{1}{2\pi}\int \textmd{d}^2p\Omega_p^z=\frac{J}{2}\textmd{sign}(\Delta).
\end{equation}
Therefore, the sign change of $\Delta$ 
corresponds to a change in the topological properties of Eq. \eqref{hamilton}.
In addition, the Hall conductivity also exhibits a step as $\Delta$ passes through zero, 
and depends on the very same topological invariant per spin and valley as
\begin{equation}
\sigma_{xy}=\frac{e^2}{h}\mathcal{C}_J=\frac{e^2}{h}\frac{J}{2}\textmd{sign}(\Delta).
\label{hallstep}
\end{equation}
States with different values of $\mathcal{C}_J$ can be regarded as belonging to distinct phases, 
similarly to the $\sigma_{xy}$ plateau phases of the integer quantum Hall effect\cite{avron}.
Note, that low energy Hamiltonians like Eq. \eqref{hamilton} usually occur pairwise (i.e. at the $K$ and $K'$ points in the 
Brillouin 
zone for graphene). Therefore the topological invariants, 
$\mathcal{C}_J$ from different valleys, 
add up to integer (not necessarily zero) Chern numbers. A $\Delta$ from spin-orbit coupling can trigger a non-zero Chern number, 
while the contribution from different valleys due to a staggered sublattice potential or bias voltage lead to a zero Chern number, 
although each valleys can have non-trivial topology with finite $C_J$.
Spin degeneracy also leads to an additional factor of 2.

\section{Quenching the gap}

We are interested in the quantum quench dynamics when the gap varies 
as $\Delta(t)=\Delta_0{t}/{\tau}$ (up to logarithmic corrections, 
as analyzed below) and $t\in [-\tau,\tau]$. According to Kibble-Zurek 
scaling~\cite{kibble,zurek}, the resulting defect (extra electron/hole on the hole/electron rich layer, respectively, equivalent to excited states in 
the upper branch in this case\cite{fischer}) density is $\rho\sim \tau^{-d\nu/(z\nu+1)}$, which leads to 
\begin{equation}
\rho\sim\left({\Delta_0}/{\tau}\right)^{1/J}.
\label{kzscaling}
\end{equation}
The matrix structure of Eq. \eqref{hamilton} allows us to connect our problem
to the Landau-Zener (LZ) dynamics~\cite{vitanov} by analysing the solution of 
\begin{gather}
i\hbar\partial_t\Psi(t)=H\Psi(t), \hspace*{5mm} \Psi(-\tau)=\Psi_-,
\label{schroedinger}
\end{gather}
where $H\Psi_\pm=E_\pm\Psi_\pm$, and the quantity of interest is
$\Psi(\tau)$. 
Considering finite temperatures amounts to change the initial condition as a combination of positive and negative energy states.
However, as long as $k_B T\ll\Delta_0$,  our results hold.
The exact solution for the diabatic transition 
probability between final ground and excited states at momentum $p$ for $\varepsilon(p)\ll \Delta_0$ 
gives for the momentum distribution of excited states in the upper branch (Fig. \ref{fancy}) and 
and the resulting total defect density 
\begin{gather}
P_p=\exp\left(-{\pi\varepsilon^2(p)\tau}/{\hbar\Delta_0}\right),
\label{momdist}\\
\label{lzscaling}
\rho=\frac{A_c}{(2\pi\hbar)^2}\int
\textmd{d}^2pP_p=\frac{A_c\Gamma(1/J)}{4J\pi\hbar^2}\left(\frac{\hbar\Delta_0}{\pi
    c_J^2\tau}\right)^{1/J} 
\end{gather}
per valley, spin and unit cell, with $A_c$ the unit cell area.
This agrees with Kibble-Zurek scaling in Eq. \eqref{kzscaling}. However, the
present approach also provides the explicit numerical prefactor for
arbitrary $J$, similarly to the quantum Ising model~\cite{dziarmaga}. 
Note that the bigger $J$, the larger (and the more insensitive to $\tau$) the resulting defect
density, on account of the larger the number of low energy
states ($\omega^{2/J}$) within an energy window $\omega$ around the Dirac 
point.

Since the number of defects from Eq. \eqref{lzscaling} progressively  increases with decreasing $\tau$, it is important to address its validity. From Ref. \onlinecite{grandi},
the borderline between a sudden and slow quench is determined from $\hbar {d}\Delta/{d}t\sim \Delta^2$, which yields 
$\tau\Delta_0\sim\hbar$. Thus, our results apply in the slow quench regime when $\tau\Delta_0>\hbar$, while the sudden quench region sets in for $\tau\Delta_0<\hbar$.

\section{Physical realization}
\subsection{Monolayer graphene}

The $J=1$ case with $c_1=v_F\approx  10^6$~m/s is realized in
MLG~\cite{castro07}, where the spinor structure encodes the two
sublattices of the honeycomb lattice. The control or even the very existence of
a gap there remains an open issue. Dirac fermions with linear band-crossing 
can alternatively be realized in optical lattices~\cite{dutta}, 
where the on-site energies of different sublattices are under control, 
allowing for the introduction of a time dependent mass gap.  
The quantum-Hall step of Eq. \eqref{hallstep} represents the hallmark of a single Dirac cone.

\subsection{Bilayer graphene with screening}

The $J=2$ case with $c_2=1/2m$ ($m\approx 0.03m_e$) coincides with the low
energy Hamiltonian of BLG~\cite{mccannbilayer} for energies below 
$t_\perp/4$, with $t_\perp\sim 0.3-0.4$~eV the interlayer hopping, and the spinor
springs from the two layers. Keeping BLG at charge neutrality by either
isolating it from the rest of the world in a perpendicular electric field, or by
using a dual-gate structure~\cite{oostinga,ZTG+09,MLS+09,tutuc,XFL+10},
a continuous change of the gate
voltage results in  closing and reopening the gap, as the density imbalance 
between the layers is inverted. Additionally, it also changes  the topological properties 
of the model as in Eq. \eqref{hallstep}.
However, screening due to electron
interactions becomes relevant in this case, and the induced gap is related to
the external potential, $U_{ext}$ as~\cite{evcastro,min} 
\begin{gather}
2\Delta=U_{ext}+\frac{e^2d\delta n}{2A_c\varepsilon_r\varepsilon_0},
\label{gapdensity}
\end{gather}
where $\delta n=\sum_p (n_{1p}-n_{2p})$ is the dimensionless density imbalance
between the two layers with $n_{ip}$ the particle density of state $p$ on the
$i$th layer. In equilibrium, to a good approximation, the induced gap is given
by~\cite{evcastro,mccannscreening} 
\begin{gather}
\Delta=\left(1+\lambda\ln\left(\frac{4t_\perp}{|U_{ext}|}\right)\right)^{-1}\frac{U_{ext}}{2}, 
\label{adiabaticpot}
\end{gather}
and the density imbalance reads
\begin{equation}
\delta n=4\rho_0\Delta\ln\left({|\Delta|}/{2t_\perp}\right),
\label{imbal}
\end{equation}
with $\lambda=e^2d\rho_0/A_c\varepsilon_r\varepsilon_0\sim 0.1-0.5$ the
dimensionless screening strength, 
$d \approx 3.3~\text{\AA}$  the interlayer distance, $\varepsilon_0$ the
permittivity of free space and $\rho_0=A_c m/2\pi\hbar^2$ the density of
states per valley and spin in the limit $\Delta \rightarrow 0$. 
For SiO$_2$/air interface, $\varepsilon_r\approx
2.5$ ($\varepsilon_r\ = 25$ for NH$_3$, $\varepsilon_r\ = 80$ for H$_2$O), 
which reduces the effects of screening. 

\section{Numerics}

In a quench of a time dependent external potential in BLG, the
induced gap couples the two-level systems (stemming from the $2\times 2$
structure of Eq. \eqref{hamilton}, labeled by $p$) via the $\delta n$
term in Eq. \eqref{gapdensity}. The problem would require the solution of a 
continuum of coupled differential equations, which is not easy, 
even approximately. 
We mention that the case of a single level (only one $p$ mode), 
in which case $\delta n=n_{1p}-n_{2p}$ in Eq. \eqref{gapdensity}, 
is known as the non-linear LZ 
model~\cite{nonlinearlz}, and  the resulting dynamics differ 
qualitatively from the conventional one, possessing nonzero transition 
probability even in the adiabatic limit for strong non-linear coupling. 

The analysis is simplified considerably by the observation that a single level
cannot have a strong impact on the dynamics of the others due to the
large number of terms in the sum for $\delta n$.
Thus, it looks natural to
replace the non-linear term by an average density imbalance, independent of the
explicit time dependence of $n_{1p}(t)-n_{2p}(t)$ for a given $p$, hence
decoupling the LZ Hamiltonians for distinct $p$'s.

When  $U_{ext}$ changes fully adiabatically, the resulting gap and density
imbalance are given by Eqs.~\eqref{adiabaticpot} and~\eqref{imbal},
respectively. For slow, nearly adiabatic temporal changes of the potential,
only a small fraction of terms in the $\delta n$ sum is expected 
to behave truly diabatically (contribution from states nearest to the gap 
edges). Thus
we assume that the gap is still given by Eq.~\eqref{adiabaticpot}, and establish
self-consistency by verifying that the resulting density imbalance satisfies 
Eq. \eqref{imbal}. Although the usage of
Eq. \eqref{adiabaticpot} simplifies the picture, it still differs from the
conventional LZ form, i.e. subleading logarithmic terms are inevitably present
albeit with a reasonably small prefactor $\lambda$. Fortunately, one can invoke
the extension of the Kibble-Zurek mechanism for non-linear quenches to estimate
the resulting defect density~\cite{barankov,sen} (note the difference between a
non-linear quench on the LZ problem~\cite{barankov,sen} and the non-linear LZ
problem~\cite{nonlinearlz}). The logarithmic terms in Eq.~\eqref{adiabaticpot} 
can be considered as "zeroth" powers, therefore the resulting quench is still 
"linear", with subleading logarithmic corrections.

\begin{figure}[h!]
\includegraphics[width=6cm,height=6cm]{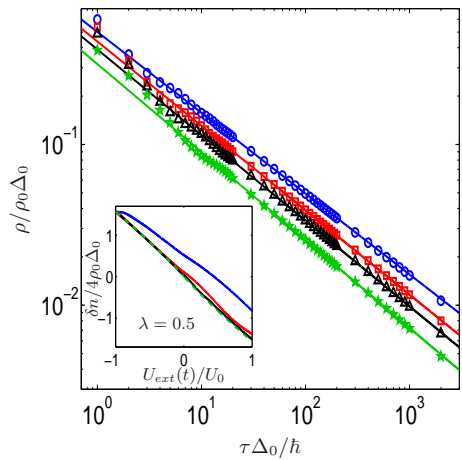}
\caption{(Color online) The density of defects created during the quench per
  spin, valley and unit cell in BLG {with} screening is shown for
  $U_{ext}=U_0t/\tau$,  $t_\perp=5U_0$, $\lambda=0$ (blue, circle), 0.1 (red,
  square), 0.2 (black, triangle) and 0.5 (green, star) from top to bottom. The
  symbols denote the numerical data, the solid lines are fits using
  $\rho/\rho_0\Delta_{0}=\frac C 2 (\frac{\hbar}{\tau\Delta_{0}})^\alpha$. The
  inset shows the time dependent density imbalance of BLG per spin and valley
  in a linear external potential {with} strong screening ($\lambda=0.5$) with
  $\tau\Delta_{0}/\hbar=1$ (blue), 10 (red) and 100 (black) from top to
  bottom. The green dashed line shows the fully adiabatic (equilibrium) result
  with $\tau\rightarrow\infty$, Eqs. \eqref{adiabaticpot}-\eqref{imbal}, which
  is approached fast with increasing $\tau$. Given the simplicity of our
  self-consistent average field procedure, the agreement is excellent for slow
  quenches.
\label{figscaling}}
\end{figure}

The inset of Fig. \ref{figscaling} shows the density imbalance, 
obtained from solving numerically the LZ problem (Eq.~\eqref{schroedinger}) 
with the adiabatic screening potential (Eq.~\eqref{adiabaticpot}) for BLG 
with a linearly varying external potential, 
\begin{equation} 
U_{ext}(t)=U_0{t}/{\tau},\hspace*{6mm} t\in [-\tau,\tau]. 
\end{equation}
The numerical results are compared to those from Eqs.~\eqref{adiabaticpot} 
and~\eqref{imbal}; the imbalance is rather well described by the
equilibrium, fully adiabatic ($\tau\rightarrow\infty$) expression (dashed-green
line), therefore our decoupling of the coupled non-linear LZ problem by the
adiabatic potential for slow enough quenches with Eq.~\eqref{adiabaticpot}
works satisfactorily. This validates our average field decoupling
procedure. 
Note that to the density imbalance in Eq. \eqref{imbal} 
all states up to the cutoff, $t_\perp$, are contributing.
On the other hand, 
defect production occurs at very low energies, close to the touching point 
of the gapless branches, whose contribution to the imbalance is negligible 
in the limit of the size of the initial gap, Eq.~\eqref{adiabaticpot},
 $\Delta_\lambda\equiv\Delta\arrowvert_{U_{ext}=U_0}\ll t_\perp$.

The number of defects (excited states in the upper branch) created in an external potential,
$U_{ext}(t)=U_0{t}/{\tau}$, $t\in [-\tau,\tau]$, follows 
Eq. \eqref{lzscaling} even in the presence of screening as
\begin{equation}
\frac{\rho}{\rho_0\Delta_{0}}=\frac 12
\sqrt{\frac{\Delta_\lambda}{\Delta_{0}}}\sqrt{\frac{\hbar}{\tau\Delta_{0}}}, 
\label{scalingscreening}
\end{equation}
where $\Delta_{0}=|U_{0}/2|$, $\Delta_\lambda\equiv\Delta\arrowvert_{U_{ext}=U_0}$. 
Eq. \eqref{scalingscreening} together with Eq. \eqref{lzscaling} are the central results of our Kibble-Zurek analysis.
The numerical data  fitted with
$\rho/\rho_0\Delta_{0}=C(\frac{\hbar}{\tau\Delta_{0}})^\alpha/2$, and both the
prefactor $C$ and the exponent $\alpha$ are compared to the expected values,
namely $\sqrt{\Delta_\lambda/\Delta_{0}}$ for the coefficient and 1/2 for the
$\tau$ exponent for various values of $\lambda$, summarized in 
Table~\ref{tab1}, and shown in Fig. \ref{figscaling}. The agreement 
is indeed remarkable, the slight mismatch in the exponent 1/2 being due to the 
subleading logarithmic terms in Eq.~\eqref{adiabaticpot} for stronger 
screening. Since $\Delta_0\rho_0\sim 10^{-3}$ for $\Delta_0\sim t_\perp/10$, 
the resulting density of defects per unit area (including spin and valley) 
falls into the order of $\sqrt{\hbar/\tau\Delta_0}\times 10^{12}$~cm$^{-2}$, 
and can take the value $3\times 10^9$~cm$^{-2}$ for quenching time 
$\tau\sim 1$~ns, corresponding to a ramping rate $\Delta_0/\tau\sim 10^7$~eV/s.
Note that this density corresponds to the electrons/holes in the otherwise 
empty/occupied upper/lower branch,
and does not by itself
imply any particular real space density modulation, since these states
contribute negligibly to the layer charge imbalance.  
A moderately slow quench implies $\tau\Delta_0/\hbar\sim 10-100$ with 
$\Delta_0\sim t_\perp/10$, translating to $\tau\sim 0.1-1$~ps. 
Different non-linear sweep protocols~\cite{barankov,sen,grandi} lead to similar
conclusion: the steeper (more non-adiabatic) the quench, the bigger the defect
density produced.

Our results are robust with respect to variations in the band structure, e.g. extra hopping terms or large
asymmetry gap.  
The quadratic spectrum of BLG with $J=2$ changes to linear one ($J=1$) at the vicinity of the Dirac point
($\sim 10$~K range) due to trigonal warping, which could affect the scaling of the defect density 
($1/\sqrt{\tau}\rightarrow 1/\tau$) for slow quenches.
Excitonic effects can either renormalize the gap in biased BLG or open small gaps in unbiased 
BLG\cite{exciton}, which can be overcome by the electric field without affecting our 
findings.


\begin{table}[h!]
\centering
\begin{tabular}{|c|c|c|c|c|}
\hline
$\lambda$ & 0 & 0.1 & 0.2 & 0.5 \\
\hline
$\sqrt{\Delta_\lambda/\Delta_{0}}$ from Eq. \eqref{adiabaticpot} & 1.00 & 0.88
& 0.80 & 0.64\\ 
\hline
 $\sqrt{\Delta_\lambda/\Delta_{0}}$ from the fit  & 1.00 & 0.87 &0.78 & 0.64\\ 
\hline
exponent ($\alpha$) & 0.50 & 0.52 & 0.53 & 0.55\\
\hline
\end{tabular}
\caption{The numerically obtained values of the coefficient,
  $\sqrt{\Delta_\lambda/\Delta_{0}}$ and the exponent 1/2 of the defect density
  from Fig. \ref{figscaling} for $t_\perp=5U_0$, compared to the values based
  on Kibble-Zurek scaling and Eq. \eqref{adiabaticpot}.}
\label{tab1}
\end{table}

\section{Population inversion, dynamic conductivity}

Having established the scaling properties of the defect density in BLG, we turn
to the determination of the optical response of the excited state resulting 
from the quench, whose momentum distribution is given by Eq. \eqref{momdist}; 
The occupation number in the upper and lower branches of the spectrum is,
respectively,  $f_+(p)=P_p$ and $f_-(p)=1-P_p$ due to particle-hole symmetry.
For momenta close to the $K$ point, population inversion occurs when 
$f_+(p)>f_-(p)$, i.e. in the energy range
$2\Delta_\lambda<\hbar\omega<2\Delta_\lambda\sqrt{1+(\hbar\ln
  2)/(\pi\Delta_\lambda\tau)}$, 
which translates in the near adiabatic limit to 
\begin{gather}
2\Delta_\lambda<\hbar\omega<2\Delta_\lambda+\frac{\hbar\ln 2}{\pi\tau}.
\label{freki}
\end{gather}
The effect of a small ac electric field can be considered using Fermi's golden
rule, and the initial dynamic conductivity  is related to the rate of optical
transitions between the two states with the same momentum,
weighted by the probabilities of occupied initial and empty final states, as
\begin{gather}
\Gamma_p(\omega)=\frac{2\pi}{\hbar}
M_p^2\delta\left(\hbar\omega-2\sqrt{\Delta_\lambda^2+\varepsilon^2(p)}\right)[f_-(p)-f_+(p)],
\label{fgr}
\end{gather}
where $M_p=|v_x(p)eA|$ is the transition matrix element between the higher and
lower energy state, where $v_x(p)=\Psi_+^* \partial H/\partial p_x\Psi_-$ 
and $A$ the vector potential. Thence, we obtain the
dynamic conductivity
\begin{gather}
\sigma(\omega)=\sigma_0
\left[1-2\exp\left(\frac{\pi\tau}{4\hbar\Delta_\lambda}(4\Delta_\lambda^2-(\hbar\omega)^2)\right)\right]\times\nonumber\\
\times\frac{(\hbar\omega)^2+4\Delta_\lambda^2}{(\hbar\omega)^2}\Theta\left(|\hbar\omega|-2\Delta_\lambda\right),
\end{gather}
with $\sigma_0=e^2/2\hbar$ the ac conductivity of
BLG~\cite{abergel,nicol}.

Both absorption and stimulated
emission are taken into account, and the negativity of the resulting 
conductivity indicates the dominance of the latter: this indicates  a phase 
coherent response, which is of course essential for a laser. In addition, stimulated 
emission can also win against spontaneous emission by increasing the intensity 
of the incoming radiation field. If spontaneous emission dominates 
(luminescence), the resulting radiation will still be spectrally limited but 
without phase coherence. 

In the frequency range of Eq. \ref{freki}, the dynamic conductivity is negative
due to the population inversion~\cite{ryzhiinegative} (i.e. the energy injected
into the system during the quench is released) as
\begin{equation}
\sigma(\hbar\omega\rightarrow 2\Delta_\lambda)\approx-2\sigma_0.
\end{equation}
 The region
of negative conductivity shrinks with increasing $\tau$, without influencing
the amplitude of $\sigma(\omega)$ precisely at the gap edge. This follows from
Eq. \eqref{momdist}, implying maximal population inversion at the Dirac point
for arbitrary quench time, i.e. $P_{p=0}=1$. 
For higher frequencies, 
$\sigma(\omega)$ is still suppressed with respect to the adiabatic optical 
response.

The typical lasing frequency lies in the
close vicinity of $\Delta_\lambda$ (including the THz regime,  wavelength of
the order of 10~$\mu$m), conveniently tunable by perpendicular electric
fields~\cite{ZTG+09}. 
The relaxation times for intra- and interband processes in MLG
are estimated as 1~ps and 1-100~ns\cite{ryzhiinegative}, respectively,  which might be further enhanced in BLG
around half-filling\cite{monteverde}. Thus, the lasing is expected to survive for 
quenching times in the ps-ns range even in the presence of the above processes.
Repeated quenching (like optical pumping) between $\Delta$ and $-\Delta$ is also 
linked to the Kibble-Zurek theory\cite{mukherjee} with similar effects on the population inversion.

The dc conductivity also reveals the effect of the electric field quench. In
the presence of a clean gap, excitations, and hence the dc conductivity, 
are exponentially suppressed at low temperatures. 
The excited electrons/holes in the upper/lower branch resulting from this 
quench, can carry a current that is not activated.

\section{Conclusions}

To conclude, by exploiting the tunability of the band gap in BLG by a perpendicular electric field, a
finite rate temporal electric field quench leads to excited state production, whose distribution is
analyzed in terms
of Kibble-Zurek scaling, LZ dynamics for non-linear quenches and is compared to the full numerical
solution of the problem with screening corrections, using an adiabatic decoupling procedure.
The effect of the quench is manifested in population inversion, and BLG could be used as a coherent
sourse of infra-red radiation, and possibly as a laser. 

Our results apply to other systems with a quadratic band crossing,
e.g. for certain nodal superconductors or  cold atoms on  Kagome or
checkerboard optical lattices~\cite{sun} at appropriate fillings, described by
Eq. \eqref{hamilton} with $J=2$ at low energies. The momentum distribution,
Eq. \eqref{momdist} and the concomitant scaling of the defect
density after closing and reopening the gap would be direct evidence of the
quench dynamics. Particularly intriguingly, graphene multilayers with
appropriate stackings realize higher order ($J>2$) band
crossings~\cite{multilayer,MM08}.

\begin{acknowledgments}
This work was supported by the Hungarian Scientific Research Fund No. K72613 and CNK80991
and by the Bolyai program of the Hungarian Academy of Sciences. 
EVC acknowledges financial support from the Juan de la Cierva Program
(MCI, Spain) and the Est\'imulo \`a Investiga\c{c}\~ao Program 
(Funda\c{c}\~ao Calouste Gulbenkian, Portugal).
\end{acknowledgments}

\bibliographystyle{apsrev}
\bibliography{refgraph}

\end{document}